\documentclass[journal=jacsat,manuscript=article,layout=traditional]{achemso}
\newif\ifjournal\journalfalse
\newif\ifarxiv\arxivtrue
\setkeys{acs}{email=false}

\setkeys{acs}{maxauthors = 0}
\usepackage{upgreek}
\usepackage{siunitx}

\usepackage[version=3]{mhchem} %
\newcommand{\oldunit}[1]{\,\mathrm{#1}}

\usepackage{color}
\definecolor{darkblue}{rgb}{0.2,0.2,0.7}

\title{Single Nitrogen--Vacancy-NMR of Amine-Functionalized Diamond Surfaces}
\author{John M. Abendroth}
\affiliation[ETH]{Department of Physics, ETH Zurich, Otto-Stern-Weg 1, 8093 Zurich, Switzerland}
\ifjournal
\email{jabendroth@phys.ethz.ch}
\fi
\alsoaffiliation[equal]{Equally contributing authors}
\author{Konstantin Herb}
\affiliation[ETH]{Department of Physics, ETH Zurich, Otto-Stern-Weg 1, 8093 Zurich, Switzerland}
\alsoaffiliation[equal]{Equally contributing authors}
\author{Erika Janitz}
\affiliation[ETH]{Department of Physics, ETH Zurich, Otto-Stern-Weg 1, 8093 Zurich, Switzerland}
\author{Tianqi Zhu}
\affiliation[ETH]{Department of Physics, ETH Zurich, Otto-Stern-Weg 1, 8093 Zurich, Switzerland}
\author{\newline Laura A. Völker}
\affiliation[ETH]{Department of Physics, ETH Zurich, Otto-Stern-Weg 1, 8093 Zurich, Switzerland}
\author{Christian L. Degen}
\affiliation[ETH]{Department of Physics, ETH Zurich, Otto-Stern-Weg 1, 8093 Zurich, Switzerland}
\ifjournal
\email{degenc@ethz.ch}
\fi

\begin{document}
\ifarxiv
\singlespacing
\fi

\subsection{Abstract}
Nuclear magnetic resonance (NMR) imaging with shallow nitrogen—vacancy (NV) centers in diamond offers an exciting route toward sensitive and localized chemical characterization at the nanoscale. Remarkable progress has been made to combat the degradation in coherence time and stability suffered by near-surface NV centers using suitable chemical surface termination. However, approaches that also enable robust control over adsorbed molecule density, orientation, and binding configuration are needed. We demonstrate a diamond surface preparation for mixed nitrogen- and oxygen-termination that simultaneously improves NV center coherence times for emitters $<$10-nm-deep and enables direct and recyclable chemical functionalization \textit{via} amine-reactive crosslinking. Using this approach, we probe single NV centers embedded in nanopillar waveguides to perform $^{19}$F NMR sensing of covalently bound trifluoromethyl tags in the \textit{ca.} 50--100 molecule regime. This work signifies an important step toward nuclear spin localization and structure interrogation at the single-molecule level.

\clearpage

Elucidating molecular structure at the single- to few-molecule level is an important goal in analytical chemistry, biochemistry, and molecular biology. It allows to study functional differences in populations that possess static disorder. As a popular structural probe, nuclear magnetic resonance (NMR) spectroscopy enables identification of atomic arrangements and hierarchical ordering in molecules using the resonant frequency of atomic nuclear spins. However, conventional NMR spectroscopy suffers from poor sensitivity due, in part, to intrinsically small nuclear spin polarizations in thermal equilibrium at room temperature.\cite{nmrsensitivity1} Alternatively, a powerful route to realize highly sensitive nanoscale-NMR employs nitrogen–vacancy (NV) centers in diamond.\cite{Mamin_science_2013,Staudacher_science_2013,Aslam_science_2017} The NV center is a fluorescent crystal defect in the diamond lattice composed of a substitutional nitrogen and adjacent vacancy that can act as an atomic-sized sensor for small magnetic moments\cite{Jelezko_PRL_2004,Schirhagl_ARPC,GaliReview}. For sensing nuclear spins at diamond surfaces, it is critical to stabilize shallow (\textless10-nm-deep) NV centers, preserve their coherence properties, and control adsorbed molecule density, orientation, and binding configuration.

\par Although functionalization of the diamond surface itself is possible using a variety of chemical attachment strategies,\cite{Hartl2004,Stavis_pnas,surfacereview2,diamondfuncreview} such modifications can be detrimental to near-surface NV centers. Charge state conversion from the useful negatively charged NV centers to the neutral state due to surface charge traps or band bending renders the defect ineffective.\cite{PhysRevB.83.081304,PhysRevLett.123.146804,PhysRevB.82.115449} Moreover, detection sensitivity is worsened in shallow NV centers from surface magnetic impurities.\cite{PhysRevB.86.081406,LoretzAPL2014,PhysRevLett.112.147602} Promising chemical approaches to minimize surface noise include fluorination,\cite{Cui_APL2013} nitrogen termination,\cite{StaceyAM,NL111N,Kawai_JPCC2019} or oxygen-termination.\cite{KavianiNL,PhysRevX.9.031052,Ohashi2013,Finkler} On oxygen-terminated diamond, for instance, carboxyl groups can then be used to attach molecules of interest using carbodiimide crosslinker chemistry.\cite{Lovchinsky_science_2016} However, immobilization density and surface passivation with adsorbed analytes are difficult to control when limited solely by the native surface density of residual chemically addressable surface groups. 

\par Atomic layer deposition (ALD) may be used to grow 1--2-nm-thick adhesion layers for dense molecular self-assembly. Recent reports have utilized this method with chemical modification by phosphate or silane anchoring groups to enable NMR detection of molecular films in ensemble NV measurements\cite{SurfaceBucher} and to develop biocompatible surface architectures for NV sensing.\cite{xie2021biocompatible} Still, this method can result in decreased coherence of near-surface NV centers, and adds distance between the NV sensor spin and target surface spins of interest. The signal-generating dipolar interaction inversely scales with the third power of this separation distance. Thus, highly sensitive nuclear spin detection with single NV centers would benefit from direct molecular attachment to the diamond surface.\cite{Kolkowitz_PRL_2012,mullerNatcomm2014,Sushkov_PRL_2014,1d2dmagresDegen,Cujia_Nature_2019} Dense molecular assemblies of silane molecules may instead be formed without adhesion layers on oxygen-terminated diamond, anchored to surface hydroxyl moieties.\cite{Notsu_2001,Ohta_2004,Silaneagain} Indeed, we have successfully performed surface NMR sensing of $^{19}$F with shallow NV centers without ALD layers for vapor-deposited films of trimethoxy(3,3,3-trifluoropropyl)silane, as well as films of (3-aminopropyl)trimethoxysilane with subsequent amine-reactive crosslinking of trifluoromethyl tags (\textbf{Supplementary Information Figure S1}). However, the mechanisms of silanization chemistry of surfaces is still not completely understood,\cite{aptms,spencer1,SiNplasma} and preventing multilayer formation or film degradation under aqueous conditions and air exposure is a major challenge.\cite{Silaneloss,VANDENBERG1991103} Improved surface treatments are necessary that simultaneously enable predictable molecular attachment and stabilization of shallow NV sensors.

\begin{figure}
\begin{center}
\includegraphics[width=5in]{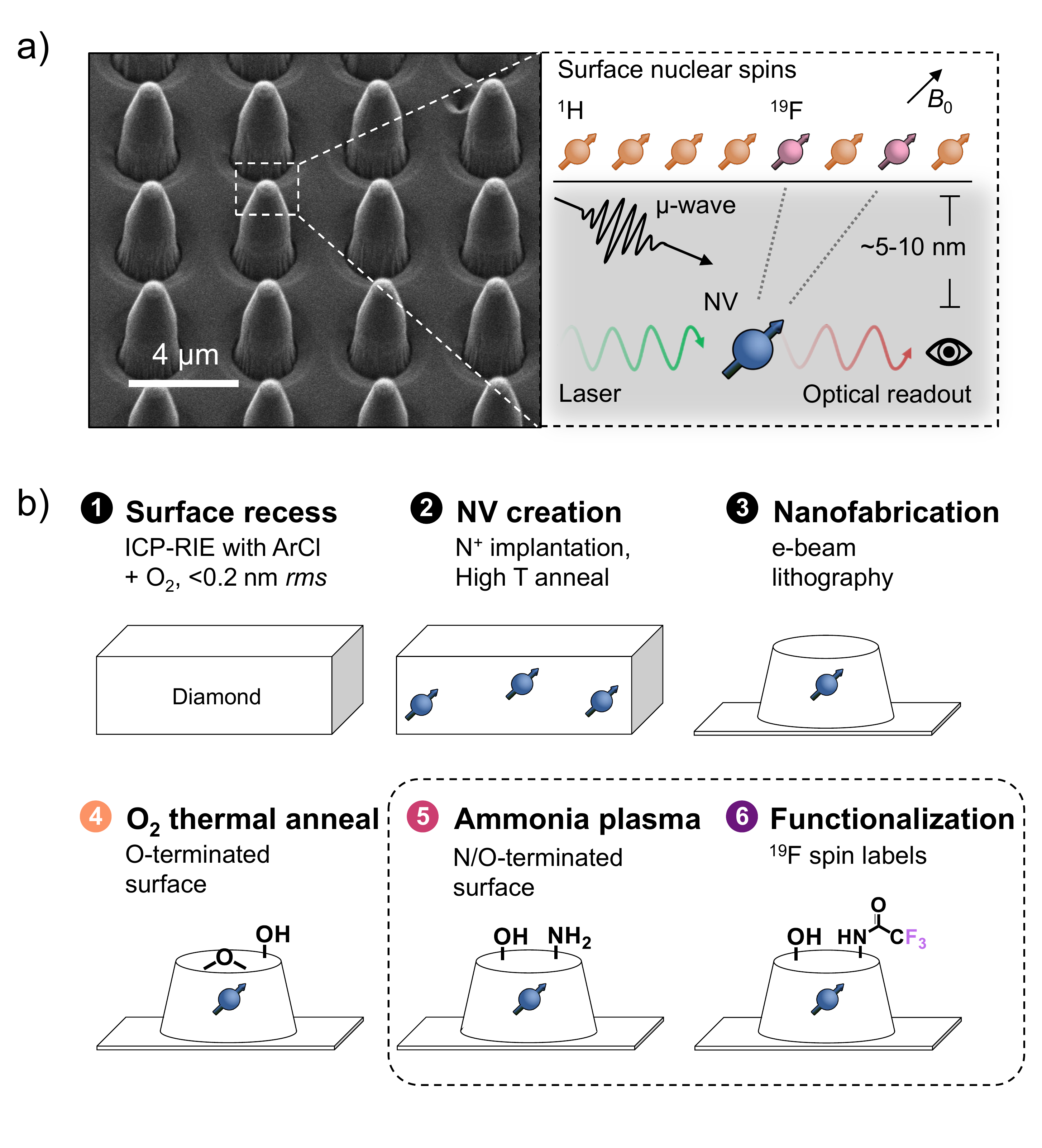}
\caption{\textbf{a)} Experimental overview for conducting surface NMR with single NV centers. Left: scanning electron micrograph of a representative diamond nanopillar waveguide array hosting NV centers. Right: scheme illustrating sensing of surface nuclear spins by shallow NV centers. \textbf{b)} Workflow of sample preparation and chemical functionalization of diamond surfaces for surface nuclear spin detection. The dotted box highlights the focus of this work.}
\label{overview}
\end{center}
\end{figure}
\clearpage
\par Here, we demonstrate direct chemical functionalization of mixed nitrogen- and oxygen- (N/O-) terminated diamond for surface NMR spectroscopy with single NV centers (\textbf{Figure~\ref{overview}}).  %
Starting with O-terminated diamond obtained through thermal annealing, the composition of mixed N/O surface termination is controlled with exposure to ammonia (NH$_3$) plasma.  We show reversible surface functionalization of amine groups introduced by plasma exposure using derivatization;  \textit{N}-hydroxysuccinimide (NHS) ester crosslinking chemistry is used to attach fluorescent dye molecules by stable amide bond formation. We find at short plasma exposure times that the mixed N/O-terminated surfaces improve the coherence time of $<$10-nm-deep NV centers compared to only O-annealed surfaces.  %
We finally functionalize diamond nanopillars hosting single NV centers with trifluoromethyl tags to detect the magnetic field variance arising from surface-bound $^{19}$F spins. Within the sensing volume of each NV, we estimate detection of \textit{ca.} 50--100 surface-bound molecules. This surface functionalization demonstrates the generalizability of our immobilization protocol, as well as the persistent stability of shallow NV centers after chemical attachment of proximal molecules for nanoscale NMR.

 Low diamond surface roughness is essential for surface sensing as it facilitates the formation of highly ordered chemical termination, which increases the coherence time of near-surface NV centers.\cite{PhysRevX.9.031052} Bulk and membrane samples (the latter for nanofabrication of pillar arrays) are recessed using Ar/Cl$_2$ and and O$_2$ inductively coupled plasma -- reactive ion etching (ICP-RIE). This step is followed by cleaning in a tri-acid mixture of H$_{2}$SO$_{4}$:HClO$_{4}$:HNO$_{3}$ at 120 \textdegree C to remove graphitic carbon on the surface as a result of the O$_2$ etching (Step 1, \textbf{Figure~\ref{overview}b}). Using this procedure we achieve surface roughness $<$0.2 nm-rms (\textbf{Supplementary Information Figure S2}). 

\par To create shallow NV centers, recessed diamond samples are implanted with $^{15}$N$^{+}$ ions with controllable density and depth (Step 2, \textbf{Figure~\ref{overview}b}). Implantation energies of 5 or 7 keV and fluences of $10^9 \oldunit{cm^{-2}}$ are used, yielding an expected average implantation depth of 8.0(31) and 10.8(40) nm respectively (\textbf{Supplementary Information Figure S3}).\cite{SRIM} Annealing at 800 \textdegree C for 2 h  under high vacuum ($<5\times 10^{-8}\oldunit{mbar}$) converts implanted $^{15}$N$^{+}$ ions into NV centers due to vacancy migration. Nanopillar arrays are then patterned on membrane samples \textit{via} electron-beam lithography (Step 3, \textbf{Figure~\ref{overview}b}) for higher photoluminescence collection efficiency and to provide a map to selectively address and revisit the same NV centers for repeated characterization.\cite{HAUSMANN2010621}

\par Samples are subsequently thermally annealed under an oxygen atmosphere at 460 \textdegree C so as to remove further graphitic carbon and to terminate the diamond surface with oxgyen (Step 4, \textbf{Figure~\ref{overview}b}). Using an analogous oxygen annealing procedure, Sangtawesin \textit{et al.} recently demonstrated improvements in coherence (\textit{T}$_2$) times of shallow NV centers of up to a factor of four compared to acid-cleaned diamond surfaces, which was attributed to a highly ordered, predominantly ether-terminated surface.\cite{PhysRevX.9.031052} Similarly, we see an improvement in NV optical contrast using both continuous wave (cw) and pulsed optically detected magnetic resonance (ODMR) measurements, and factor of two improvement in coherence and Rabi decay times of NV centers following oxygen annealing (\textbf{Supplementary Information Figure S4}). %

\par Although suggested to be beneficial for near-surface NV stability, the resulting scarcity of hydroxyl or carboxylic acid functional groups on O-terminated surfaces by thermal annealing in oxygen precludes straightforward functionalization with molecules of interest. Here, we incorporate additional reactive amine terminal groups for subsequent molecular attachment by introducing a 50 W \textit{rf} of 13.56 MHz NH$_3$ plasma treatment to O-annealed surfaces for mixed N/O-termination (Step 5, \textbf{Figure~\ref{overview}b}). This approach mitigates full conversion to a surface that exhibits negative electron affinity (characteristic of fully amine-terminated diamond),\cite{ZHU2016295} which can be detrimental for NV stability. We note that alternative diamond amination procedures using NH$_3$, N$_2$, or mixed-source plasmas have been developed previously, but with hydrogen-terminated diamond as a starting surface or without characterizing their influence on the photophysical properties of near-surface NV centers.\cite{nh3plaspeptide,WEI201534,OxygenNH3role,wang_langmuir_2012,chandran_APL} %

\begin{figure}
\begin{center}
\includegraphics[width=5in]{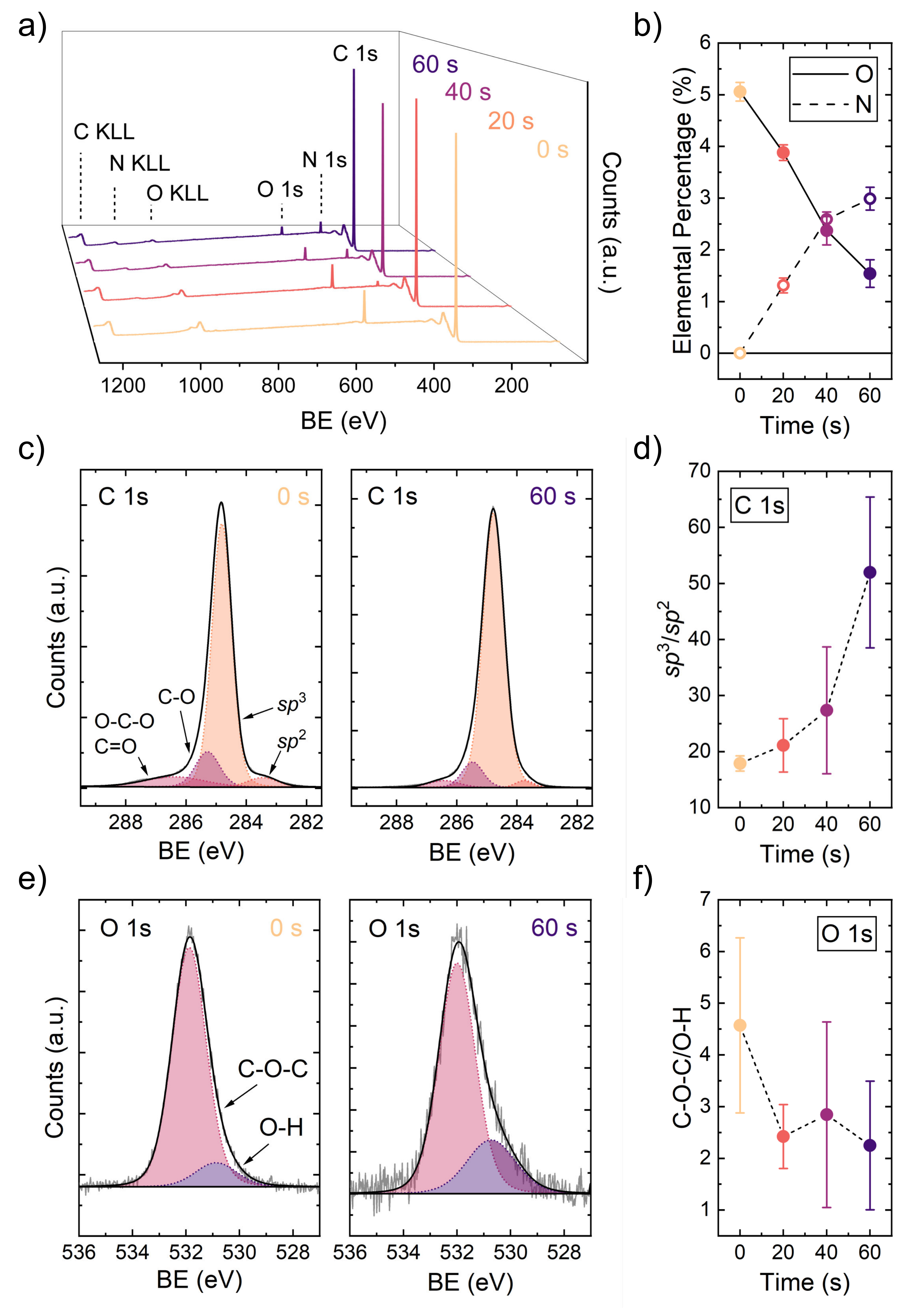}

\end{center}
\end{figure}

\begin{figure}
\begin{center}

\caption{Characterization of diamond surfaces with X-ray photoelectron spectroscopy. \textbf{a)} Representative survey scans of oxygen-annealed diamond surfaces with increasing exposure time to NH$_3$ plasma. Binding energies (BE) of C, O, and N 1s core electrons and Auger signatures are indicated. \textbf{b)} Elemental percentage of oxygen and nitrogen within the sampling depth as a function of exposure time. \textbf{c)} Representative high-resolution C 1s scans before (0 s) and after (60 s) plasma exposure. \textbf{d)} Ratio of \textit{sp}$^3$/\textit{sp}$^2$ components from the fitting of C 1s signals. \textbf{e)} Representative high-resolution O 1s scans before (0 s) and after (60 s) plasma exposure. \textbf{f)}  Ratio of C--O--C (ether) assignment at 531.9 eV and O--H (hydroxyl) contribution at \textit{ca.} 1 eV lower binding energy from fitting of O 1s signals. Error bars represent standard deviation from \textit{N} = 3 samples for each condition.}
\label{xpsfig}
\end{center}
\end{figure}

\par Changes to the chemical composition of the surfaces exposed to 0, 20, 40, and 60 s of NH$_3$ plasma are characterized using X-ray photoelectron spectroscopy (XPS) on bulk oxygen-annealed diamond samples. A decrease in elemental percentage of oxygen in the sampling depth (\textit{ca.} $\le$6 nm) is accompanied by an increase in nitrogen to 3\% after 60 s (\textbf{Figure~\ref{xpsfig}a,b}). These percentages are obtained by integration of O 1s, N 1s, and C 1s signals after adjusting for photoionization cross sections of the core electron species.\cite{YEH19851} The emergent N 1s peak position at 398.4 eV suggests predominately single-bonded C--N and/or double-bonded C=N, while nitrile groups may be excluded.\cite{GRAF20092849,StaceyAM} Analysis of high-resolution C 1s spectra shows a dominant peak at 284.8 assigned to \textit{sp}$^3$ carbon, a shoulder at \textit{ca.} 1 eV lower binding energy attributed to \textit{sp}$^2$ carbon, and satellite peaks at higher binding energy due to carbon--oxygen binding (\textbf{Figure~\ref{xpsfig}c}).\cite{PhysRevX.9.031052,xps1} The NH$_3$ plasma treatment reduces the already low amount of \textit{sp}$^2$ carbon on the diamond surfaces with increasing exposure time (\textbf{Figure~\ref{xpsfig}d}). The high-resolution O 1s signals can be fit with a C--O--C (ether) assignment at 531.9 eV and C--O--H (hydroxyl) contribution at \textit{ca.} 1 eV lower binding energy (\textbf{Figure~\ref{xpsfig}e}).\cite{PhysRevX.9.031052} The relative ratio of these two peaks for oxygen-annealed diamond is \textit{ca.} 5:1, and decreases to \textit{ca.} 2:1 after 60 s plasma exposure (\textbf{Figure~\ref{xpsfig}f}). 

\begin{figure}
\begin{center}
\includegraphics[width=5in]{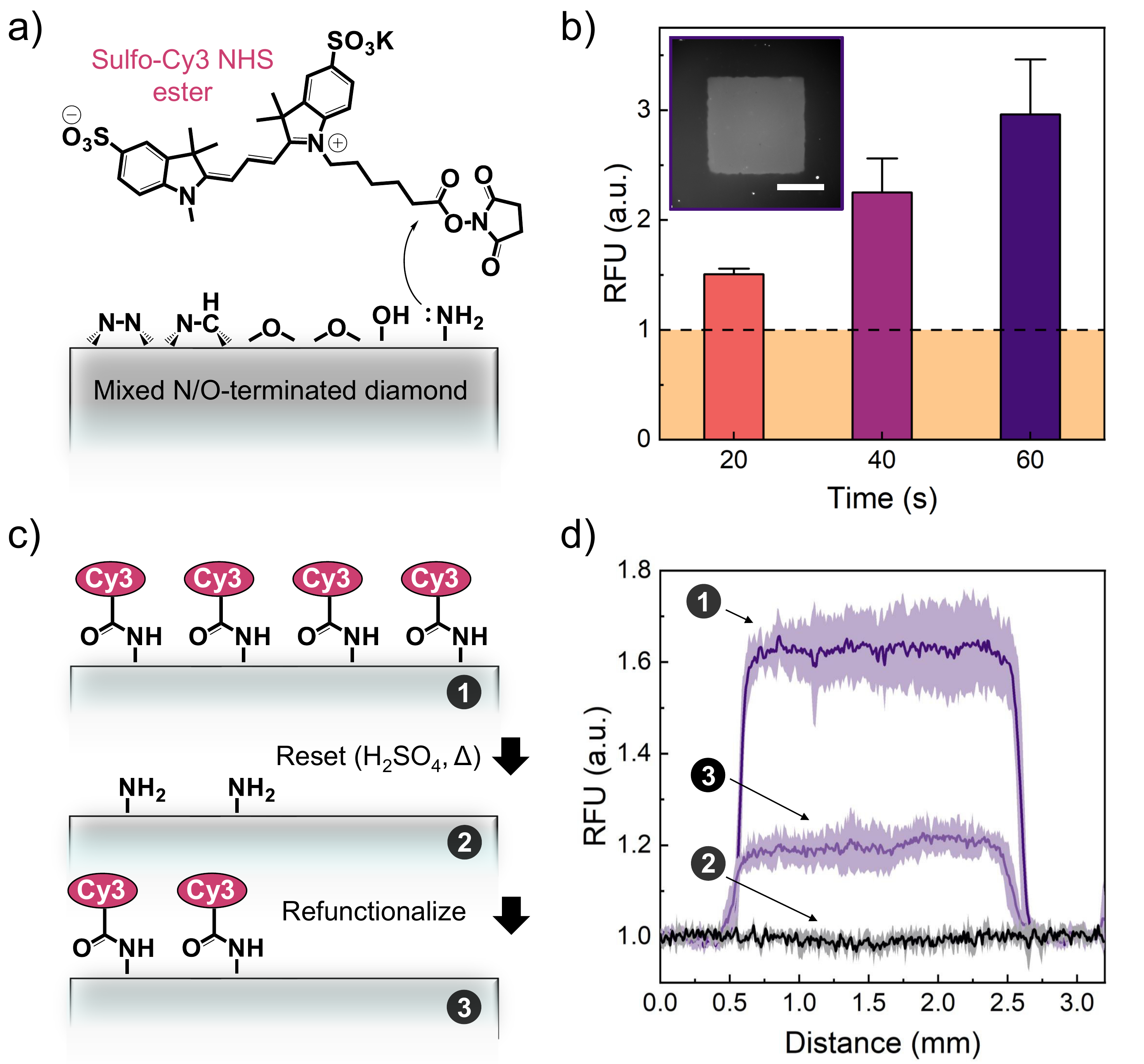}
\caption{Characterizing diamond surface functionalization \textit{via} fluorescent dye attachment. \textbf{a)} Schematic of amine derivatization with the fluorescent dye, sulfo-Cy3 NHS ester, on mixed N/O-terminated surfaces. \textbf{b)} Fluorescence intensity measured as relative fluorescence units (RFU) of patterned square regions exposed to NH$_3$ plasma over oxygen-only-terminated background regions ($\mathrm{RFU}=1$). \textit{N} = 5 per condition. Error bars represent standard errors of the means. Inset shows a representative fluorescence microscopy image of a dye-functionalized square region exposed to 60 s plasma (scale bar, 0.2 mm). \textbf{c)} Schematic of surface recyclability. \textbf{d)} Fluorescence linescans (average of \textit{N} = 5 per condition) from the same exposed square region on a diamond surface after sulfo-Cy3 NHS ester functionalization (1), after resetting the surface with sulfuric acid (2), and after a second dye functionalization step (3). Errors represent standard deviation.}
\label{fluoro}
\end{center}
\end{figure}

The presence of reactive amines on the mixed N/O-terminated diamond surface is determined using complementary fluorescence microscopy characterization (\textbf{Figure~\ref{fluoro}a,b}). Amine groups are derivitized by incubating surfaces with a sulfo-Cy3 NHS ester dye designed to bind selectively \textit{via} amine-reactive crosslinking. From 20 s to 60 s exposure time, increasing relative fluorescence intensity is observed above the O-terminated background regions. These results indicate clearly the tunability of available amine groups and surface density of adsorbed molecules. This functionalization is also reversible (\textbf{Figure~\ref{fluoro}c,d}). We show that surfaces can be reset by treating with 75\% H$_2$SO$_4$ at 60 \textdegree C for 5 h in order to break the amide bonds formed between the adsorbed molecules and aminated surface, indicated by total loss of fluorescence relative to the background region. A repeated functionalization of the surface by incubating with sulfo-Cy3 NHS ester again shows fluorescence from the region originally exposed to NH$_3$ plasma, albeit at lower intensity. The reduction in intensity is likely due to partial loss of surface-bound amine groups following the reset, resulting in a lower surface density of dye molecules. Still, this proof-of-principle recyclability may be optimized to preserve a higher percentage of reactive functional groups for repeated analysis.

We next test how the NH$_3$ plasma treatment affects the sensing properties of shallow NV centers. \textbf{Figure~\ref{bulkNVdata}a} shows representative confocal fluorescence images of bulk diamond surfaces containing individually resolvable NV centers before and after 20 s plasma exposure. The plasma treatment causes background fluorescence to decrease, making it easier to distinguish fluorescent NV centers from bright surface impurities. Next, individual NV centers are identified and characterized using OMDR, wherein the NV center is first illuminated with a green laser ($\lambda=532$ nm) to polarize the electron spin into the $m_{s}=0$ state. Applying a subsequent microwave field at the Larmor frequency of the electron spin drives transitions between the optically polarized $m_{s}=0$ (bright) state and the $m_{s}=\pm$1 (dark) states. Finally, readout is performed by integrating the emitted photons during a second green laser pulse. The measurement sensitivity scales linearly with ODMR contrast, which is defined as the ratio of integrated fluorescence of the bright \textit{vs} dark states \cite{Schirhagl_ARPC}. %

 We find that increased plasma exposure time is accompanied by a decrease in the number of NV centers that display an ODMR signal (\textbf{Figure~\ref{bulkNVdata}b}). The survival rate is calculated by determining the number of NV centers with an ODMR signal detected per scan area after plasma treatment, and normalized to the number of detected NV centers prior to exposure (\textbf{Supporting Information Figure S5}). Approximately two-thirds of NV centers remain following 20 s plasma exposure, which drops below 50\% after 40 and 60 s. Consequently, we opted to use a nanostructured diamond membrane for further testing such that we could track the same NV centers before and after treatment. \textbf{Figure~\ref{bulkNVdata}c} shows a representative fluorescence map of one nanopillar array after a 20 s plasma treatment. Analogous to the bulk samples, a decrease in fluorescence background is observed, in addition to a decrease in the number of pillars containing NV candidates with detectable ODMR signals. Encouragingly, no significant difference in the mean pulsed-ODMR contrast nor the mean photoluminescence of NV centers is determined before \textit{vs} after 20 s plasma exposure (\textbf{Figure~\ref{bulkNVdata}d,e}).  %

\begin{figure*}
\begin{center}
\includegraphics[width=5in]{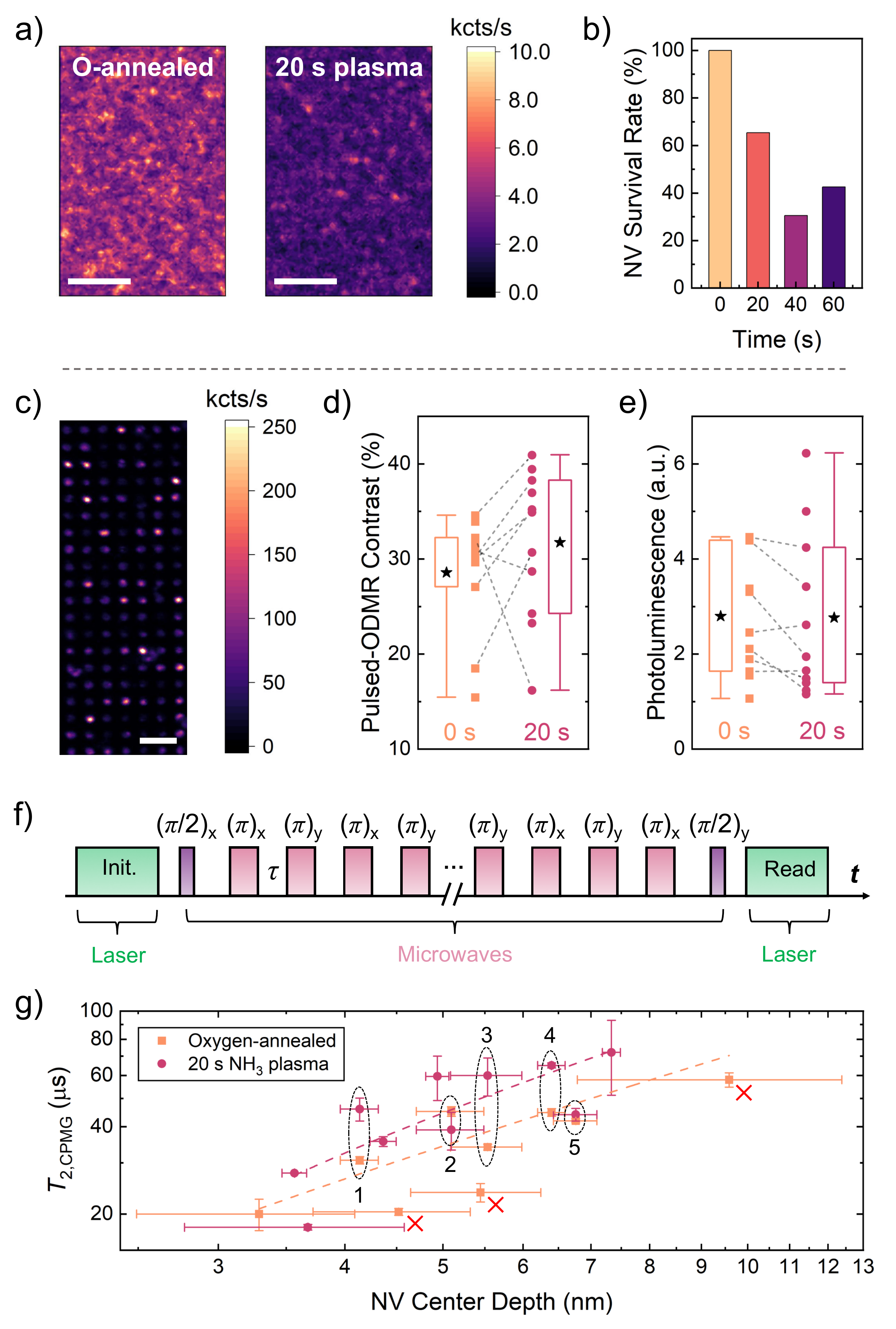}
\end{center}
\end{figure*}
\begin{figure}
\begin{center}
\caption{Comparison of NV center properties before and after exposure to NH$_3$ plasma. \textbf{a)} Representative confocal fluorescence image of a bulk diamond surface containing shallow NV centers before plasma exposure (left) and after 20 s (right). Scale bars are 5 $\si{\micro m}$. \textbf{b)} Survival rate of NV centers after different plasma exposure times. \textbf{c)} Representative confocal fluorescence image of a nanopillar array hosting NV centers in a diamond membrane. Scale bar is 10 $\si{\micro m}$.  \textbf{d)} Pulsed-ODMR contrast and \textbf{e)} photoluminescence intensity of NV centers in nanopillar waveguides before and after plasma treatment. In the box and whisker plots, black stars show the mean values of the data. Boxes compose the middle 25\%--75\% of the data. Whiskers extend to 1.5$\times$ this interquartile range. Dotted grey lines connect the same NV centers characterized before and after plasma exposure. \textbf{f)} Dynamical decoupling spectroscopy sequence for detecting nuclear spin signals. \textbf{g)} Coherence time (\textit{T}$_{2},_{\mathrm{CPMG}}$) of NV centers as a function of depth within the diamond. Red crosses indicate NV centers from which no ODMR signal could be measured following plasma treatment. Dashed lines are guides for the eye. Data points representing the same NV centers characterized both before and after plasma exposure are indexed. \textbf{a)} and \textbf{b)} correspond to data collected from bulk crystals, while \textbf{c)}--\textbf{g)} correspond to data collected from nanopillar arrays on diamond membranes.}
\label{bulkNVdata}
\end{center}
\end{figure}

{We compare the coherence properties of NV centers using a Carr-Purcell-Meiboom-Gill-type (CPMG) multi-pulse dynamical decoupling protocol,\cite{CarrPurcell,MeiboomGill} which is illustrated in \textbf{Figure~\ref{bulkNVdata}f}. After an initial $\pi/2$ pulse, a train of $\pi$ pulses follows which refocuses the spin. A final $\pi/2$ pulse maps the coherence back to the initialized bright state. When increasing the evolution time between the $\pi$ pulses, we observe a signal decay with which we associate the $T_{2,\mathrm{CPMG}}$ time. To compensate for brightness variations, we collect reference traces of an initialization of the $m_S=0$ and $m_s=-1$ state in parallel. The sequence also allows us to perform NMR spectroscopy. By tuning the inter-pulse delay of the $\pi$ pulses a CPMG sequence, we match the nuclear Larmor frequency and detect the noise generated by external nuclear spins in a lock-in manner.\cite{Degen_Review} By quantifying the signal, we estimate the root mean square value of the $B$ field fluctuations produced by external spins. This signal can then be used to infer the depth of the NV center. Fit analysis providing $B_\mathrm{rms}(^1\mathrm{H})$ is performed by a Monte-Carlo model as well as by fitting an analytical model. \cite{LoretzAPL2014,PhysRevB.93.045425,Herb2022} From the extracted $B_\mathrm{rms}$ we infer the depth by inverting\cite{LoretzAPL2014}
\begin{equation}
\begin{split}
    B_\mathrm{rms}^2(d) &= \frac{5 \mu_0^2 \hbar^2 \gamma_\mathrm{H1}^2}{1536 \pi} \frac{\rho}{d^3} \left( 1 - \frac{d^3}{(d+\Delta)^3} \right) \\ &\approx (1.14\,\mathrm{\mu T\, nm^3})^2 \frac{\rho}{d^3} \left(1-\frac{d^3}{(d+\Delta)^3}\right),
\end{split}
\end{equation}
where $\mu_0$ denotes the magnetic field constant, $\hbar$ is the reduced Planck's constant, $\gamma_\mathrm{H1}=2 \pi \cdot  (42.58\,\mathrm{MHz/T})$ is the gyromagnetic ratio of protons, $\rho$ is the density of the surface proton layer, $\Delta$ is the thickness of this layer and $d$ is the NV center depth. We assume an adsorbed water layer of 1 nm thickness and proton density of $\rho=60\,\mathrm{nm}^{-3}$.\cite{LoretzAPL2014,Degen_PNAS_2009,Mamin_NL_2009,Xue_PRB_2011,Grob_NL_2019} 

The \textit{T}$_{2,\mathrm{CPMG}}$ coherence times are shown as a function of NV center depth in \textbf{Figure~\ref{bulkNVdata}g}. We observe a monotonic decrease in NV center coherence time with increasing proximity to the surface, consistent with previous reports.\cite{Myers_PRL_2014,Oliveira_NComm2017,PhysRevX.9.031052,nesladek1,xie2021biocompatible} Promisingly, the \textit{T}$_{2,\mathrm{CPMG}}$ values are larger on average after 20 s plasma exposure. In analysis of five NV centers directly compared before and after plasma treatment, we observe increases in \textit{T}$_{2,\mathrm{CPMG}}$ by as much as 1.8(3) times. Depths and \textit{T}$_{2,\mathrm{CPMG}}$ values of these NV centers are summarized in \textbf{Table \ref{tab:t2}}. In addition, from the data set characterized prior to plasma exposure, we could not measure an ODMR signal from three of the NV centers following 20 s plasma, indicated by red crosses in \textbf{Figure~\ref{bulkNVdata}g}. We note that these NV centers, with calculated depths of \textit{ca.} 4.5(8), 5.4(8), and 9.6(28) nm, each exhibited below-average coherence times, suggesting that while the plasma reduces the number of observed NV centers (\textbf{Figure~\ref{bulkNVdata}b}), it preserves those with the highest stability and sensitivity for performing surface NV-NMR measurements.}

\begin{table}[]
    \centering
    \begin{tabular}{l c||c|c c}
        { }& Depth\textsuperscript{a} $d$ (nm) & \multicolumn{2}{c}{\textit{T}$_{2,\mathrm{CPMG}}$ ($\oldunit{\upmu s}$)}  & {}\\
        { }&  & O-annealed & NH$_3$ plasma & {}\\
        \cline{2-4}
         NV 1 & 4.1(2) & 30.6(9) & 46.0(42) & {} \\
         NV 2 & 5.1(4) & 45.1(17) & 39.0(58) & {} \\
         NV 3 & 5.5(4) & 34.0(8) & 60.0(9) & {} \\
         NV 4 & 6.4(2) & 44.8(14) & 65.0(13)  & {}\\
         NV 5 & 6.8(3) & 41.9(7) & 44.0(22) & {} \\
         \multicolumn{5}{l}{\small\textsuperscript{a} Depth values are calculated from the average of}\\
         \multicolumn{5}{l}{\small  $B_\mathrm{rms}$ measurements before and after plasma treatment.}
\end{tabular}
    \caption{Depth\textsuperscript{a} and \textit{T}$_{2,\mathrm{CPMG}}$ for NV centers characterized before and after 20 s NH$_3$ plasma exposure.}
    \label{tab:t2}
\end{table}

\begin{figure}
\begin{center}
\includegraphics[width=5in]{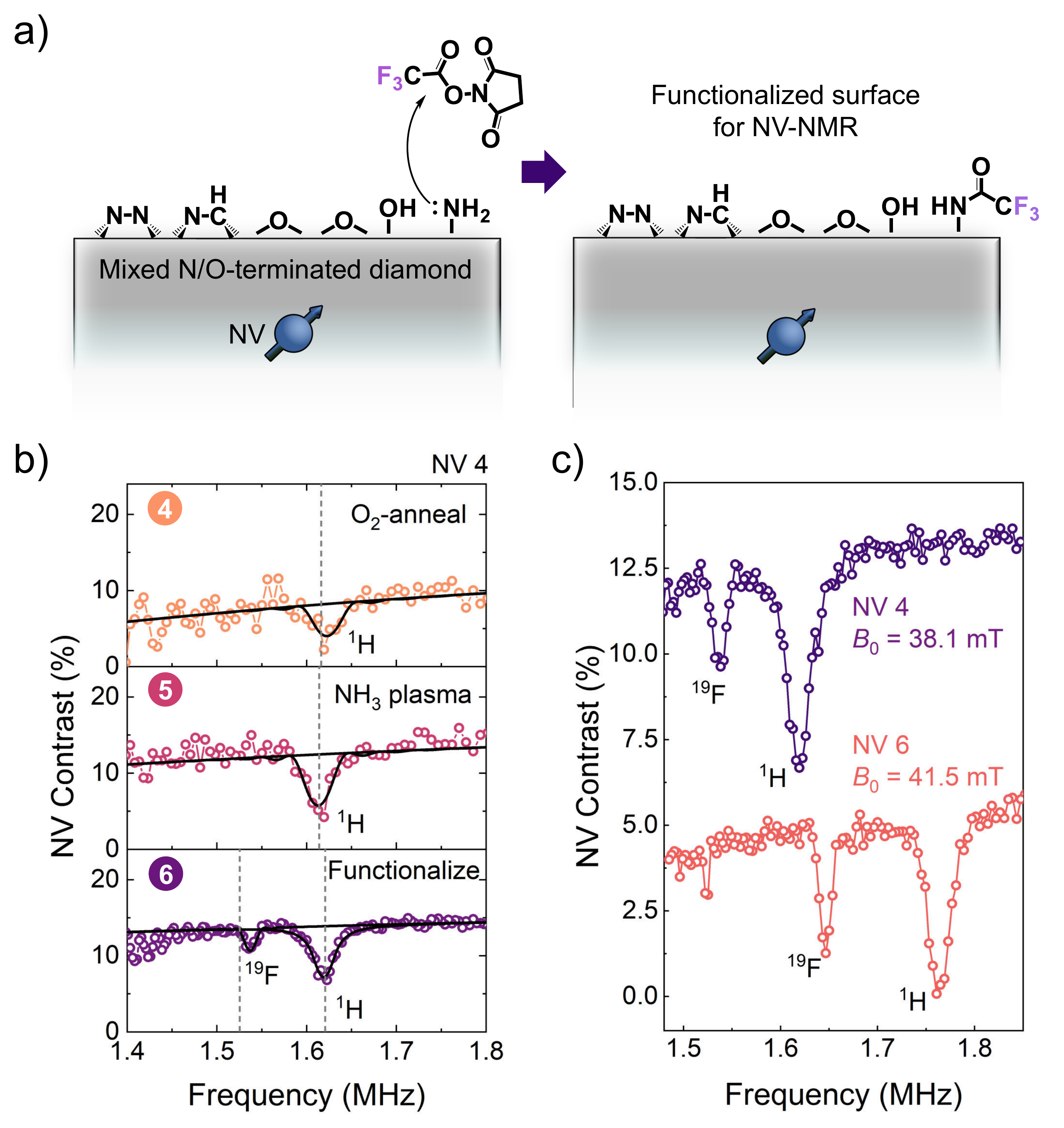}
\caption{NV-NMR sensing of surface nuclear spins. \textbf{a)} Schematic of surface functionalization with 2,5-dioxopyrrolidin-1-yl 2,2,2-trifluoroacetate. Amine groups on mixed N/O-terminated diamond surfaces serve as anchoring points to attach $^{19}\mathrm{F}$ nuclear spin labels. \textbf{b)} Representative $^1\mathrm{H}$ and $^{19}\mathrm{F}$ NV-NMR data from the same NV center during surface treatment. Top: after oxygen annealing. Middle: after 20 s ammonia plasma exposure. Bottom: after functionalization with $^{19}\mathrm{F}$ nuclear spin labels. Dotted vertical lines indicate expected positions of $^{19}\mathrm{F}$ and $^{1}\mathrm{H}$ resonance frequencies. \textbf{c)} NV-NMR data collected at two bias fields $B_0$. Top and bottom traces (NV 4 and NV 6) were recorded using two different NV centers. For the top trace, we find the $^{1}\mathrm{H}$ resonance at 1619(1) kHz, the $^{19}\mathrm{F}$ resonance at 1537(1) kHz. For the bottom trace at 1764(1) kHz and 1647(1) kHz, respectively.}
\label{nmrfig}
\end{center}
\end{figure}

{Finally, we test our N/O-termined diamond NV sensing platform for nanoscale-NMR of surface-bound molecules (Step 6, \textbf{Figure~\ref{overview}b}). Following plasma treatment, a diamond nanopillar sample is incubated with 2,5-dioxopyrrolidin-1-yl 2,2,2-trifluoroacetate to attach trifluoromethyl groups to the surface \textit{via} amide bond formation (\textbf{Figure~\ref{nmrfig}a}). Functionalization is confirmed by detection of fluorine on the surface \textit{via} XPS.} 

Surface NV-NMR spectra are shown in \textbf{Figure~\ref{nmrfig}b} that correspond to the \textit{same} NV center (NV 4) measured after oxygen-annealing, 20 s plasma treatment, and surface functionalization (Steps 4--6, \textbf{Figure~\ref{overview}b}). Prior to attachment of trifluoromethyl moieties to the surface, the dynamical decoupling spectrum shows only dips in contrast at the frequency corresponding to $^{1}\mathrm{H}$ spins, attributed mainly to the presence of an adsorbed water layer. After exposure to 20 s plasma, an improvement in coherence time of the NV is evident by reduced signal decay with decreasing frequency and a further pronounced proton signal due to this slower decay.  Following attachment of trifluoromethyl tags, a second dip appears in the vicinity of the expected $^{19}\mathrm{F}$ resonance frequency. Assignment in the NV-NMR spectra is validated by measuring two separate NV centers (NV 4 and NV 6) independently and at distinct magnetic field biases, which shifts the Larmor frequencies of both $^{1}\mathrm{H}$ and $^{19}\mathrm{F}$ spins (\textbf{Figure~\ref{nmrfig}c}). We find for NV 4 at the $^{1}\mathrm{H}$ frequency a $B_\mathrm{rms}$ of 245(5) nT and 102(5) nT at the $^{19}\mathrm{F}$ frequency.
For NV 6, we find 331(12) nT and 163(7) nT for these two resonances, respectively. From the proton signals, we estimate the NV depths to be 6.4(2) nm and 6.3(2) nm (Equation (1)).\cite{LoretzAPL2014,Degen_PNAS_2009,Mamin_NL_2009,Xue_PRB_2011,Grob_NL_2019} Modeling the $^{19}\mathrm{F}$ spin bath as a 2D layer, 
\begin{equation}
    B_\mathrm{rms}^2 =\frac{5 \mu_0^2 (\gamma_\mathrm{F19} \hbar )^2}{512\pi d^4}\rho_\mathrm{2D} = (1.86\,\mathrm{\mu T nm^{3}})^2 \frac{\rho_\mathrm{2D}}{d^4},
\end{equation}
where $\rho_\mathrm{2D}$ is the density of the 2D layer and $\gamma_\mathrm{F19}=2\pi\cdot(40.078\,\mathrm{MHz/T})$ is the gyromagnetic ratio of $^{19}\mathrm{F}$. 
By using the depths determined by $^1\mathrm{H}$ NMR, we estimate the surface density of $^{19}\mathrm{F}$ to be 5(1) $^{19}\mathrm{F}$ $\mathrm{nm}^{-2}$ and 12(2) $^{19}\mathrm{F}$ $\mathrm{nm}^{-2}$ for NV 4 and NV 6, respectively. Since each attached sub-unit contains three fluorine atoms per molecule, this corresponds to a molecular surface density of 1.7(5) and 4.0(1) molecules  $\mathrm{nm}^{-2}$. By using the sensitive surface area for a NV center at depth $d$ which produces first 70\% of the signal, $A = 0.735 d^2$ (\textbf{Supporting Information Figure S6}), we conclude that 51(15) and 117(10) molecules contribute to the detected signal by each NV center, respectively.\cite{LoretzAPL2014}

In summary, we demonstrate a diamond surface preparation strategy that enables direct chemical functionalization of amine functional groups while improving coherence times of near-surface NV centers for surface NMR spectroscopy. Mixed N/O-terminated surfaces are prepared by thermal annealing under an oxygen atmosphere at elevated temperature and subsequent exposure to NH$_3$ plasma. This process yields robust and reproducible control over the elemental percentages of nitrogen and oxygen on diamond surfaces. We identify an experimental regime in which two-thirds of shallow NV centers are preserved following NH$_3$ plasma treatment with improved sensing properties compared to oxygen-annealed surfaces. In particular, we find significant reduction in the background luminescence and up to 1.8(3) times improvement in the coherence times of near-surface NV centers. The use of nanostructured diamond allowed us to compare the same NV centers at different stages of processing, with the additional benefit of improved collection efficiency. Finally, we bind trifluoromethyl tags to surface-bound amine groups and perform NMR spectroscopy at the few-molecule (\textit{ca.} 50--100 molecules) regime using dynamical-decoupling noise spectroscopy with single NV centers. The presented approach enables a precise preparation of molecular systems of interest at diamond surfaces that may be applied for emergent NV center-based quantum sensing of chemical functionality.\cite{nesladek2,pnasDNA,CappellaroCOVID,Mzyk} 

\clearpage
\section{Supporting Information}

The Supporting Information provides additional materials and experimental methods, as well as additional data used to evaluate the conclusions in the paper; Figures S1--S7 and Table S1 show data related to XPS characterization and NV-NMR with silanized surfaces, SRIM (The  Stopping  and  Range  of  Ions  in Matter) calculations, atomic force microscopy images, characterization of NV centers following thermal annealing under oxygen, additional characterization of NV centers following exposure to ammonia plasma, integration of the sensitive slice on the surface above each NV center for NMR measurements, and estimated errors from magnetic field measurements and NV center depths. 

\section{Author Contributions}
J.M.A. and K.H. contributed equally to this work. J.M.A., K.H., and C.L.D. conceived and designed the experiments. T.Z. developed the nano-structure fabrication. Data were collected by J.M.A., K.H., and E.J. All authors discussed the results. The manuscript was co-written by J.M.A. and K.H. with assistance by E.J., T.Z., L.A.V., and C.L.D.
\section{Notes}
The authors declare no competing financial interest.

\section{Acknowledgment}
The authors thank Dr. Andrea Arcifa from EMPA, and Dr. Viraj Damle and Dr. Jan Rhensius from QZabre AG for insightful discussions and their help. This work has been supported by Swiss National Science Foundation (SNSF) Project Grant No. 200020 175600, the National Center of Competence in Research in Quantum Science and Technology (NCCR QSIT), and the Advancing Science and TEchnology thRough dIamond Quantum Sensing (ASTERIQS) program, Grant No. 820394, of the European Commission. J.M.A. acknowledges funding from an ETH Zurich Career Seed Grant and from a SNSF Ambizione Grant [PZ00P2\_201590].

\clearpage
\bibliography{db}

\end{document}


\ifarxiv
\singlespacing
\fi
\pagebreak
\section{Materials}
Bulk quantum-grade $\langle 100\rangle$ CVD grown single crystal diamonds (2 mm $\times$ 2 mm $\times$ 0.5 mm) were purchased from Element Six (UK) Ltd. (Didcot, UK). Trimethoxy(3,3,3-trifluoropropyl)\-silane (TMTFS), 
water (for Chromatographie (LC-MS) LiChrosolv), sulfuric acid (H$_{2}$SO$_{4}$), perchloric acid (HClO$_{4}$), nitric acid (HNO$_{3}$), hydrogen peroxide (H$_{2}$O$_{2}$) and dimethyl sulfoxide (DMSO) were purchased from Sigma-Aldrich (St. Louis, MO, USA). 2,5-Dioxopyrrolidin-1-yl 2,2,2-trifluoroacetate (DPTFA) was purchased from AmBeed (Arlington Heights, IL, USA). Sulfo-Cyanine3 NHS ester was purchased from Lumiprobe (Hanover, DE). Isopropyl alcohol (IPA) and (3-Aminopropyl)-trimethoxysilane (APTMS) were purchased from Fisher Scientific (Hampton, NH, USA).  

\section{Methods}
\subsection{Diamond Sample Preparation} Experiments were performed on both bulk single-crystal diamond samples as well as membranes which were cut from bulk crystals to \textit{ca}. 20-$\si{\micro \m}$-thick membranes and polished (Almax easyLab bvba; Diksmuide, Belgium). As-received diamond samples were first cleaned in piranha solution (3:1 H$_{2}$SO$_{4}$:H$_{2}$O$_{2}$), followed by cleaning for \textit{ca}. 4 h in a 1:1:1 tri-acid mixture of H$_{2}$SO$_{4}$:HClO$_{4}$:HNO$_{3}$ at 120 \textdegree C. The membranes were then recessed by inductively coupled plasma--reactive-ion etching (ICP--RIE) (Oxford Instruments PlasmaPro 100; Abingdon, UK) with ArCl and O$_{2}$. The samples were exposed to Ar/Cl$_2$ ICP for 1 h and 3$\times$ cycled etching with Ar/Cl$_2$ for 5 min and O$_2$ for 10 min. Tri-acid cleaning was then repeated to remove graphitic carbon on the surface as a result of the O$_2$ etching. 

For NV creation, recessed bulk samples and membranes were implanted with $^{15}$N$^{+}$ ions (CuttingEdge Ions, LLC; Anaheim, CA, USA) using implantation energies of 7 or 5 keV and fluences of $10^9 \unitold{cm^{-2}}$ which yields an expected average implantation depth of 10.8(40) and 8.0(31) nm, respectively (as determined using The Stopping and Range of Ions in Matter (SRIM\cite{SRIM}) simulations, \textit{c.f.} \textbf{Figure \ref{fig:srim}}). Samples were subsequently annealed at 800 \textdegree C for 2 h under vacuum, and finally tri-acid cleaned once more before nanofabrication steps. Nanopillar waveguide arrays were fabricated on membrane samples using electron-beam lithography and RIE etching commercially by QZabre AG (Zurich, CH). The fabricated nano-pillars are typically 300 nm in diameter and 2 $\si{\micro m}$ in height. Pillars are fabricated in 20 $\times$ 20 patches and are typically 4--5 $\si{\micro m}$ spaced within one array (\textit{c.f.} \textbf{Figure 1} of the main text).\newline
\clearpage
\subsection{Mixed N/O Diamond Surface Termination}
Diamond samples were baked at 460 \textdegree C under oxygen atmosphere (AS-Micro, RTP-System, Annealsys; Montpellier, FR) for 4 h to etch away any residual graphitic carbon and to increase the percentage of oxygen termination on the surfaces. To nitrogenate the surface, membranes were exposed to 50 W \textit{rf} of 13.56 MHz plasma with 20 sccm NH$_3$ with total pressure in the chamber of 300 mTorr at 80 \textdegree C for 20, 40, or 60 s (Oxford Instruments PECVD 80+, Abingdon, UK).\newline

\subsection{Surface Characterization by X-Ray Photoelectron Spectroscopy}
X-ray photoelectron spectroscopy was performed using a PHI Quantera SXM photoelectron spectrometer at the Swiss Federal Laboratories for Materials Science and Technology (EMPA, Dübendorf, ZH). A monochromatic Al K$\alpha$ X-ray source with a 100 $\si{\micro m}$ circular spot size was used under ultrahigh vacuum ($1\times10^{-9}$ mbar). High-resolution C 1s spectra were acquired at a pass energy of 55 eV using a 20 ms dwell time. For all scans, 15 kV was applied with an emission current of 3 mA; an average of 6--10 scans were collected over three distinct regions for each measurement condition. Spectra were analyzed with CasaXPS Software Version 2.3.23PR1.0. A Tougaard background and Gaussian-Lorenzian peak shapes were fit to integrate signals from high-resolution C 1s, O 1s, and N 1s spectra, and to deconvolute contributions to the high resolution C 1s and O 1s signals. In C 1s spectra, the dominant peaks were assigned to \textit{sp}$^3$ carbon and calibrated to 284.8 eV as a charge reference.
\newline

\subsection{Surface Functionalization}
Functionalization of mixed N/O-terminated diamond surfaces was performed \textit{via} amine-reactive crosslinking. For characterization by fluorescence microscopy, mixed N/O-terminated diamond surfaces were incubated with 10 mM sulfo-Cyanine3 NHS ester dissolved in water for 2 h to attach the fluorescent dye to the surface using NHS ester crosslinking. Surfaces were then rinsed thoroughly with water and IPA and blown dry with nitrogen. For surface NMR measurements, surfaces were incubated for 4 h with 10 mM DPTFA dissolved in a 1:9 ratio of DMSO:H$_{2}$O to attach -CF$_{3}$ moieties to terminal amine groups. Surfaces were then rinsed thoroughly with water and IPA and blown dry with nitrogen.\newline %

\subsection{Fluorescence Microscopy of Dye-Functionalized Diamond Surfaces}
Fluorescence microscopy was conducted in the The Scientific Center for Optical and Electron Microscopy (ScopeM) at ETH Zurich. Bulk diamond samples without fabricated nanopillar arrays were used to test surface functionalization \textit{via} fluorescence microscopy. Oxygen-annealed samples were treated with 20, 40, or 60 s ammonia plasma using a custom-made alumina mask to expose only $0.4\times 0.4$ mm square regions prior to surface functionalization with sulfo-Cy NHS ester. Fluorescence images were collected using a Nikon Ti2-E inverted fluorescence microscope and Orca Fusion BT camera (2304 $\times$ 2304 pixels, 6.5 $\si{\micro m}$ $\times$ 6.5 $\si{\micro m}$). To compare relative fluorescence intensity at plasma exposure times of 20, 40, and 60 s, images were collected using a CFI SPlan Fluor ELWD 20 $\times$ C (NA 0.45 Ph1 WD 8.2--6.9 mm) objective lens. To compare relative fluorescence intensity from the same exposed square regions after treatment with 60 s ammonia plasma, after surface resetting, and after refunctionalizyation, a CFI Plan Apochromat 4$\times$ (NA 0.2 WD 20 mm) objective lens was used. All images were collected with 500 ms exposure and excitation and emission wavelengths of 554/23 nm and 596/30 nm, respectively. Image processing and analysis was performed using ImageJ \cite{Schneider.2012}. Relative fluorescence units were calculated as pixel intensities inside $2\times 2$ mm square regions (mixed N/O-terminated) exposed to plasma, and normalized to pixel intensities outside of the square regions (O-terminated background). \newline

\subsection{Confocal Microscope for NV Measurements and Microwave Instrumentation}
Experiments were performed using a custom-built confocal microscope equipped with a green $\lambda=532\unitold{nm}$ frequency-doubled Nd:YAG excitation laser (CNI Laser MSL-FN-532nm) and a $630-800\unitold{nm}$ detection path using a single-photon avalanche photodiode (Pelkin Elmer SPCM-AQR Series). Optical pulses were generated by an acousto-optic modulator (AOM, Crystal Technology 3200-144) in a double-pass configuration. Gating of arriving photons was realized by time-tagging (NI-PCIe-6363) and software binning of photon counts. Counts were integrated typically for 270 to 380 ns. Typical laser excitation powers were on the order of $500\,\si{\micro W}$ for bulk samples and $100\,\si{\micro W}$ for nano-patterned samples.
Microwave pulses for manipulating the electronic spin were synthesized using an arbitrary waveform generator (AWG, Tektronix AWG5014C) and up-converted them using a vector signal generator (Stanford Research Systems SG-386) \textit{via} IQ modulation. Pulses were subsequently amplified using a broad-band linear amplifier (Amplifier Research 80S1G40). The microwave field at the NV center location origins from a home-build coplanar waveguide (CPW) photo-lithographically defined on a quartz cover-slip and reaches field strengths of typically 700 -- 1400 $\si{\micro T}$. The transmission line was terminated by using an external 50$\,\Omega$ load. A cylindrical samarium-cobalt permanent magnet (TC-SmCo, reversible temperature coefficient $0.001\%\,K^{-1}$) was used to apply the spin-aligning $B_0$ field ($B_0\sim 40\,\mathrm{mT}$). The bias field $B_0$ was aligned with the NV symmetry axis by adjusting the relative location of the permanent magnet and observing the resonances of ODMR scans performed with different magnet positions.\newline

\subsection{Characterization of NV Centers}
Measurements are run in an automated manner. After the manual pre-selection of NV candidates by identifying highly fluorescent spots in confocal scans, we conduct continuous-wave optically detected magnetic resonance (cw-OMDR) experiments to obtain the resonance frequency.  of the electron spin transition. To be able to compare also cw-ODMR properties, we ensure that measurements are taken at comparable microwave field strength and laser powers. Subsequently, we perform pulsed experiments for sites full-filling contrast thresholds ($>10\%$) and count-rate thresholds ($>150\,\mathrm{kcts/s}$). We determine the pulsed contrast and the $T_{2,\mathrm{CPMG}}$ time by fitting the background decay w.r.t the simultaneously recorded references traces for the bright and the dark state. We use for the CPMG-like sequences the phase cycles XY8 as well as XY16. The typical number of decoupling pulses applied was 96 -- 384. \newline

\subsection{{NV Center Depth Calculation}}
In order to estimate the depth of the NV centers, we quantify the spin noise produced by a proton layer on top of the diamond.
\subsubsection{Estimation of $B_\mathrm{rms}$}
From the publication of Pham et al. \cite{PhysRevB.93.045425} , we fit relation (1) in combination with relation (3) to the experimental data using lmfit.\cite{lmfit} We confirmed that the obtained $B_\mathrm{rms}$ values coincidence with Monte-Carlo simulations used before in the group to estimate the depth.  

\subsubsection{Translating $B_\mathrm{rms}$ to depth}
After obtaining the $B_\mathrm{rms}$ value by fitting, we estimate the expected noise level of a modeled water layer covering the diamond surface.
The dipolar coupling energy between the NV center spin and the bath of classical magnetic dipoles is given by 
\begin{equation}
    E_\mathrm{dipole} = \frac{\mu_0}{4\pi} \mu_I \left(\frac{3 (\vec{e}_I\cdot \vec{r})(\vec{e}_{NV}\cdot \vec{e}_r)}{r^3}-\frac{\vec{e}_I\cdot\vec{e}_{NV}}{r^3}\right)
\end{equation}
The squared magnetic field amplitude of the fluctuations of a single nuclear spin on the NV center axis for precessing in the plane perpendicular to the NV axis can be expressed as 
\begin{equation}
\begin{split}
    \hat{B}_\mathrm{rms}^2 = \left(\frac{\mu_0}{4\pi}\mu_I\right)^2 \frac{9}{\left(x'^2+y'^2+z'^2\right)^5} ((x'\sin\theta_0+z'\cos\theta_0)^2 y'^2+\\(x'\sin\theta_0+z'\cos\theta_0)^2(x'\cos\theta_0-z'\sin\theta_0)^2) \text{,}
\end{split}
\label{eq:Brmsdelta}
\end{equation}
where $\theta_0$ denotes the angle of 54.7\textdegree \, between the NV center axis and the axis of the surface normal of a $\langle 100\rangle$ cut diamond.
For an adsorbate layer of finite thickness, we integrate this expression to
\begin{equation}
\begin{split}
    B_\mathrm{rms}^2 &= \int_{x'=-\infty}^{+\infty}\int_{y'=-\infty}^{+\infty}\int_{z'=d}^{d+\Delta} \mathrm{d}x' \mathrm{d}y' \mathrm{d}z'\hat{B}_\mathrm{rms}^2 \rho_\mathrm{3D} \\&=  \int_{z'=d}^{d+\Delta} \mathrm{d}z'  \frac{5 \mu_0^2 (\gamma_I \hbar)^2}{512\pi z'^4}\rho_\mathrm{3D} = \frac{5 \mu_0^2 \hbar^2 \gamma_I^2}{1536 \pi} \frac{\rho}{d^3} \underbrace{\left( 1 - \frac{d^3}{(d+\Delta)^3} \right)}_{\rightarrow 1 \:\mathrm{for}\: \Delta\rightarrow\infty}
\end{split}
\label{eq:Brmslayer}
\end{equation}
\clearpage
Here, we use $\mu_I=\hbar \gamma_I/2$ for nuclear spins with I=1/2 and $\gamma_I$ in units of $\mathrm{rad}\,\mathrm{s}^{-1}\,\mathrm{T}^{-1}$. For $\gamma_I$ in $\mathrm{Hz}\,\mathrm{T}^{-1}$ replace $\hbar$ with $h$. 
To find the depth, we numerically invert relation \eqref{eq:Brmslayer} and solve only for real and positive solutions of $d$. 
The factor under-braced in \eqref{eq:Brmslayer} shows the correction for a layer of finite thickness $\Delta$. The influence of it for estimating the thickness is plotted in \textbf{Figure \ref{fig:scaling_thickness}a}.

\subsection{{NV-NMR and Surface Density Calculation}}
For a 2D layer of surface spins, we integrate relation \eqref{eq:Brmsdelta} only for $x'$ and $y'$ and evaluate at $z'=d$. We find
\begin{equation}
    B_\mathrm{rms}^2 =\frac{5 \mu_0^2 (\gamma_I \hbar)^2}{512\pi d^4}\rho_\mathrm{2D}
\end{equation}
By using the previously calculated depth $d$ of the NV center based on the proton NMR, we estimate the surface density $\rho_\mathrm{2D}$. Note that a small error in the inferred depth can have tremendous implications for the estimated density due to the $d^{-4}$ dependence. The determined depth is itself dependent on the assumed water layer thickness (\textit{c.f.} \textbf{Figure~\ref{fig:scaling_thickness}b}). \textbf{Table \ref{tab:thicknessvariation}} reports the estimated surface densities for various thickness assumptions. \newline

\subsubsection{Sensitive surface area}
The sensitivity profile from \eqref{eq:Brmsdelta} is plotted in \textbf{Figure \ref{fig:sensing_profile}}. We numerically integrate and find the region that contributes 50\% to $B_\mathrm{rms}^2$ or 70\% to $B_\mathrm{rms}$. This area is estimated to $A=0.735\,d^2$. 

For the 3D case, the corresponding sensitive volume is $V=(0.98 d)^3$.

\begin{figure}[H]
\begin{center}
\includegraphics[width=6in]{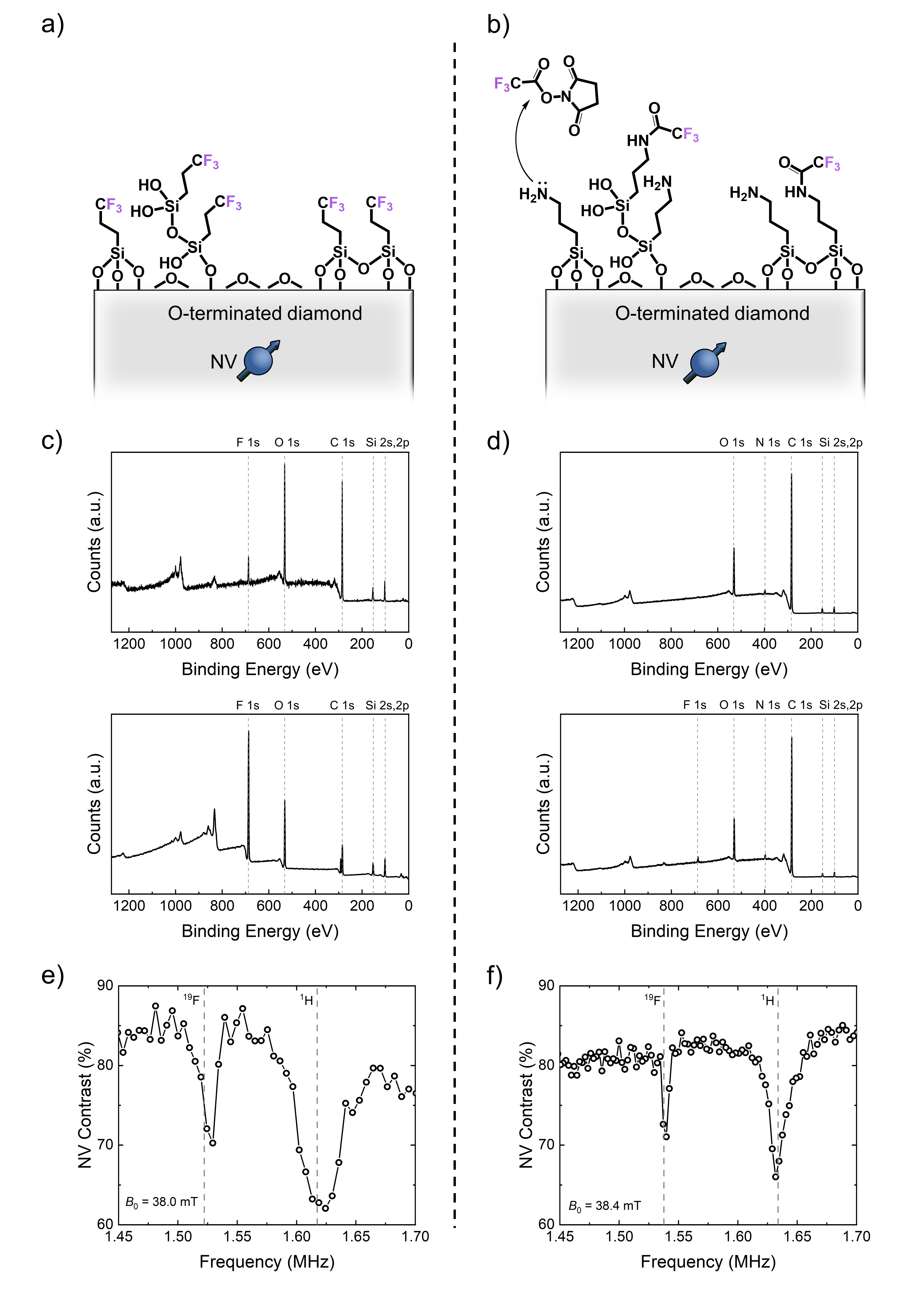}
\end{center}
\end{figure}

\begin{figure}[H]
\begin{center}
\caption{Surface functionalization \textit{via} silanization and characterization using X-ray photoelectron spectroscopy (XPS) and NV-NMR. \textbf{a)} Schematic showing oxygen-terminated diamond surfaces functionalized with $^{19}$F labeled trimethoxy(3,3,3-trifluoropropyl)silane (TMTFS) or \textbf{b)} (3-aminopropyl)-trimethoxysilane (APTMS)  which could subsequently react with 2,5-dioxopyrrolidin-1-yl 2,2,2-trifluoroacetate (DPTFA) \textit{via} amine-reactive crosslinking. Bulk samples (for XPS) or membrane samples with nanopillar arrays (for NV-NMR measurements) were baked at 460 \textdegree C under oxygen atmosphere for 4 h. Oxygen-annealed diamond samples were then exposed to thermally evaporated TMTFS or APTMS under vacuum for 1 h at 40 \textdegree C before baking under vacuum at 100 \textdegree C for 12 h to attempt removal of non-specifically adsorbed species. To subsequently attach -CF$_{3}$ moieties to terminal amine groups on diamond surfaces \textit{via} amine-reactive crosslinking, APTMS-functionalized diamond samples were incubated for 4 h with 10 mM DPTFA dissolved in a 1:9 ratio of DMSO:H$_{2}$O. Samples were rinsed thoroughly with water and IPA and blown dry with nitrogen. \textbf{c)} Two representative XPS survey scans (top and bottom) of diamond samples functionalized \textit{via} vapor deposition of TMTFS showing the high variability in F 1s signal strength. These results are indicative of poor control over multilayer formation with TMTFS. \textbf{d)} Representative XPS survey scans of diamond samples functionalized with APTMS (top) and subsequently modified by with DPTFA (bottom), which resulted in the emergence of a F 1s signal. Compared to vapor deposition of TMTFS, formation of multilayer films with APTMS appears less prevalent in our experiments. In \textbf{c) and d)}, binding energies of F 1s, O 1s, N 1s, C 1s, and Si 2s,2p core electron species are indicated. \textbf{e)} NV-NMR sensing of surface nuclear spins with single NV cetners using diamond nanopillar waveguides with surfaces functionalized with TMTFS or \textbf{f)} APTMS and DPTFA. NV-NMR traces show detection of both $^1$H and $^{19}$F nuclear spin species as indicated. NV characterizations were perfomed using the same methodology as for the other data.}
\label{fig:silane}
\end{center}
\end{figure}
\pagebreak

\begin{figure}[H]
\begin{center}
\includegraphics[width=\linewidth]{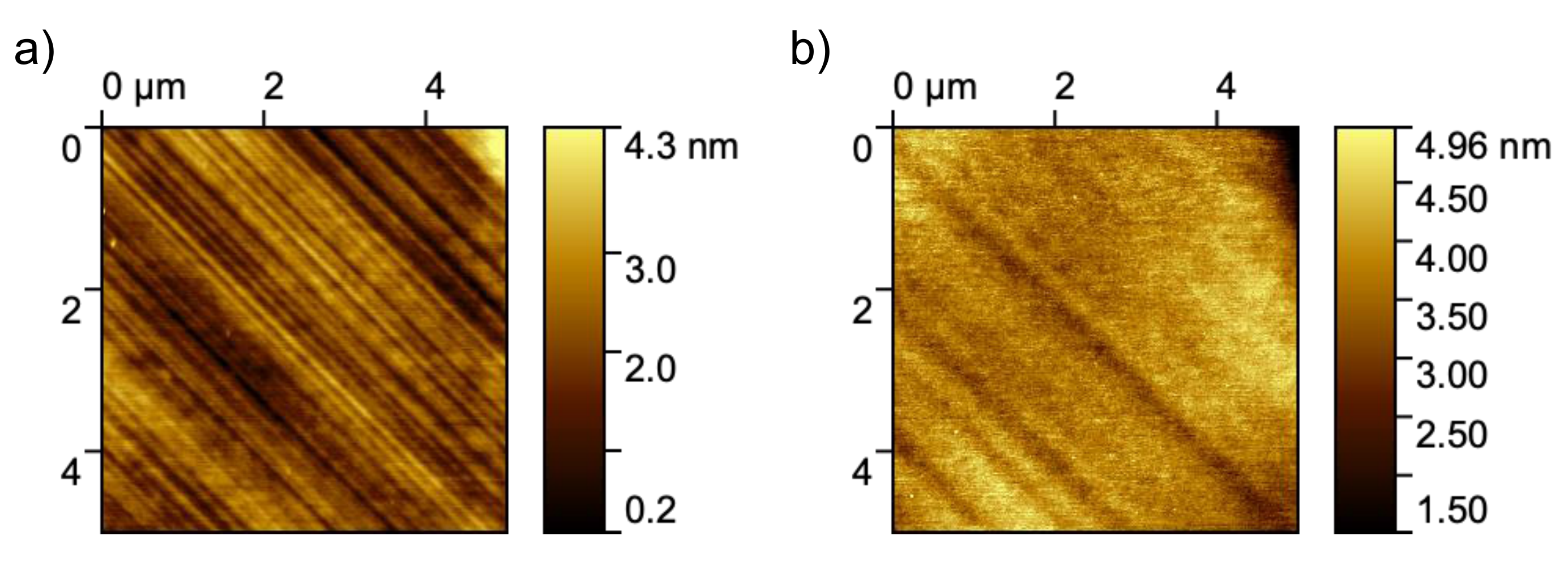}
\caption{Atomic force microscope (AFM) images of an electronic grade diamond \textbf{a)} after tri-acid cleaning, before nanofabrication, and \textbf{b)} after 1 hour of ArCl$_2$ ICP RIE followed by three cycled etches comprising 5 min of ArCl$_2$ and 10 min of O$_2$. The total etch procedure removes 5.5 $\si{\micro m}$ of material to visibly remove damage from mechanical polishing and smooth the surface.}
\label{fig:afm}
\end{center}
\end{figure}
\clearpage

\begin{figure}[H]
    \centering
    \includegraphics[width=5.5in]{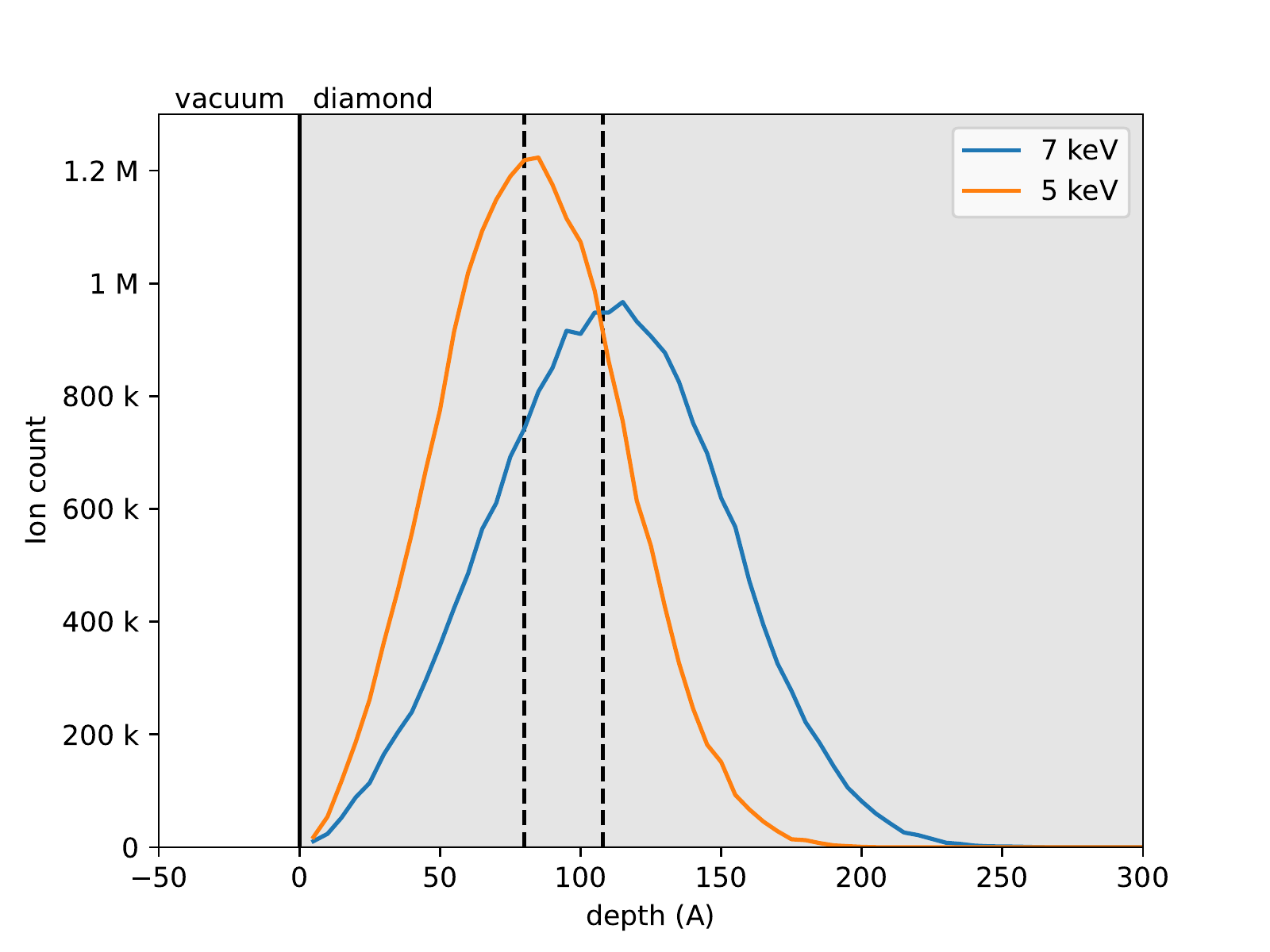}
    \caption{Stopping range calculations for $^{15}\mathrm{N}^{+}$ ion implantation as determined using The Stopping and Range of Ions in Matter, SRIM simulations.\cite{SRIM}}
    \label{fig:srim}
\end{figure}

\begin{figure}[H]
\begin{center}
\includegraphics[width=6.5in]{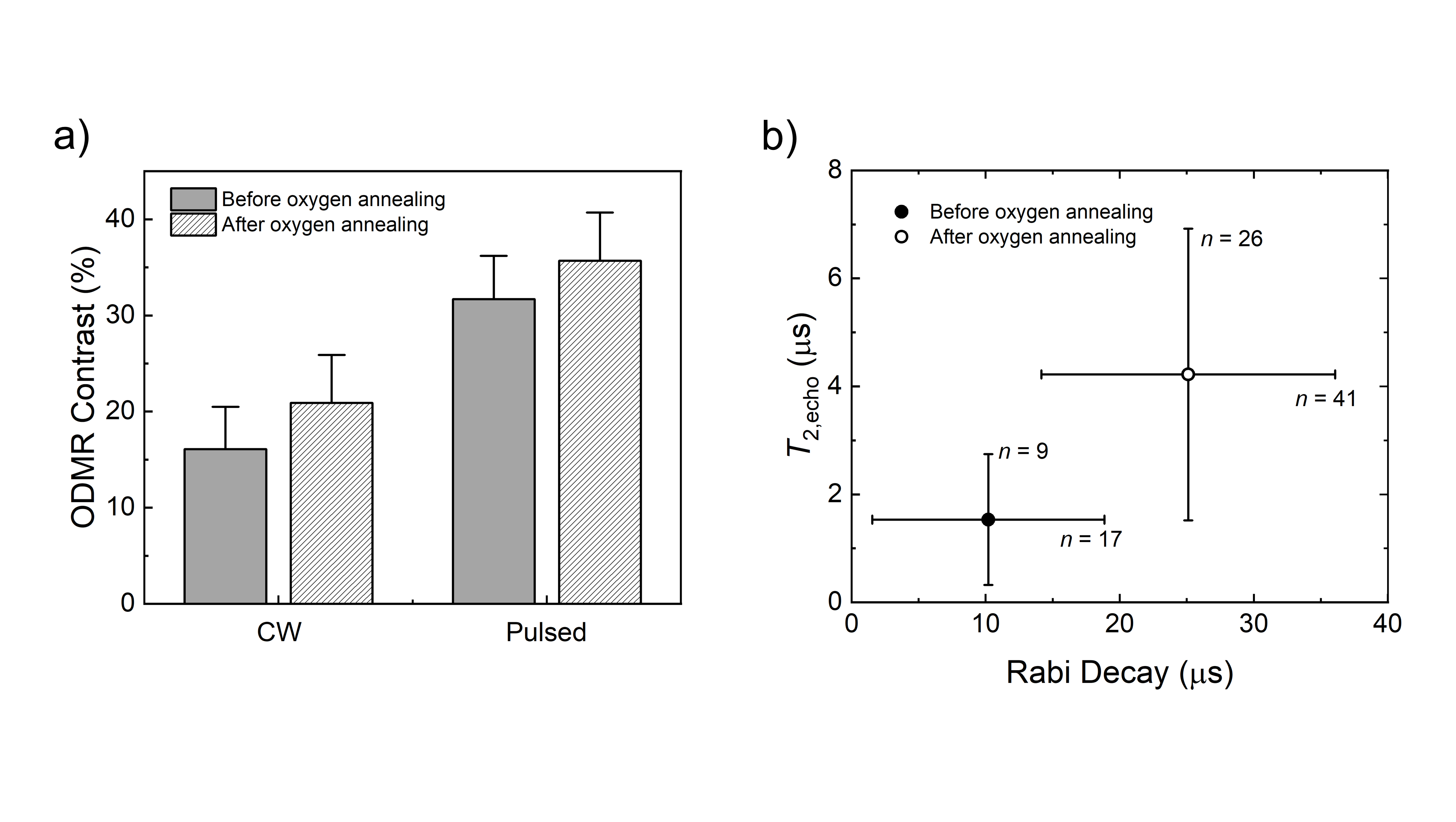}
\caption{Effect of the oxygen annealing (Step 4 of \textbf{Figure 1b} in the main text) on the contrast of the NV centers under investigation and onto the coherence time. We observe a positive effect by incorporating an O$_2$ thermal annealing treatment at elevated temperature after tri-acid cleaning.}
\label{fig:oxygen}
\end{center}
\end{figure}

\begin{figure}[H]
\begin{center}
\includegraphics[width=6in]{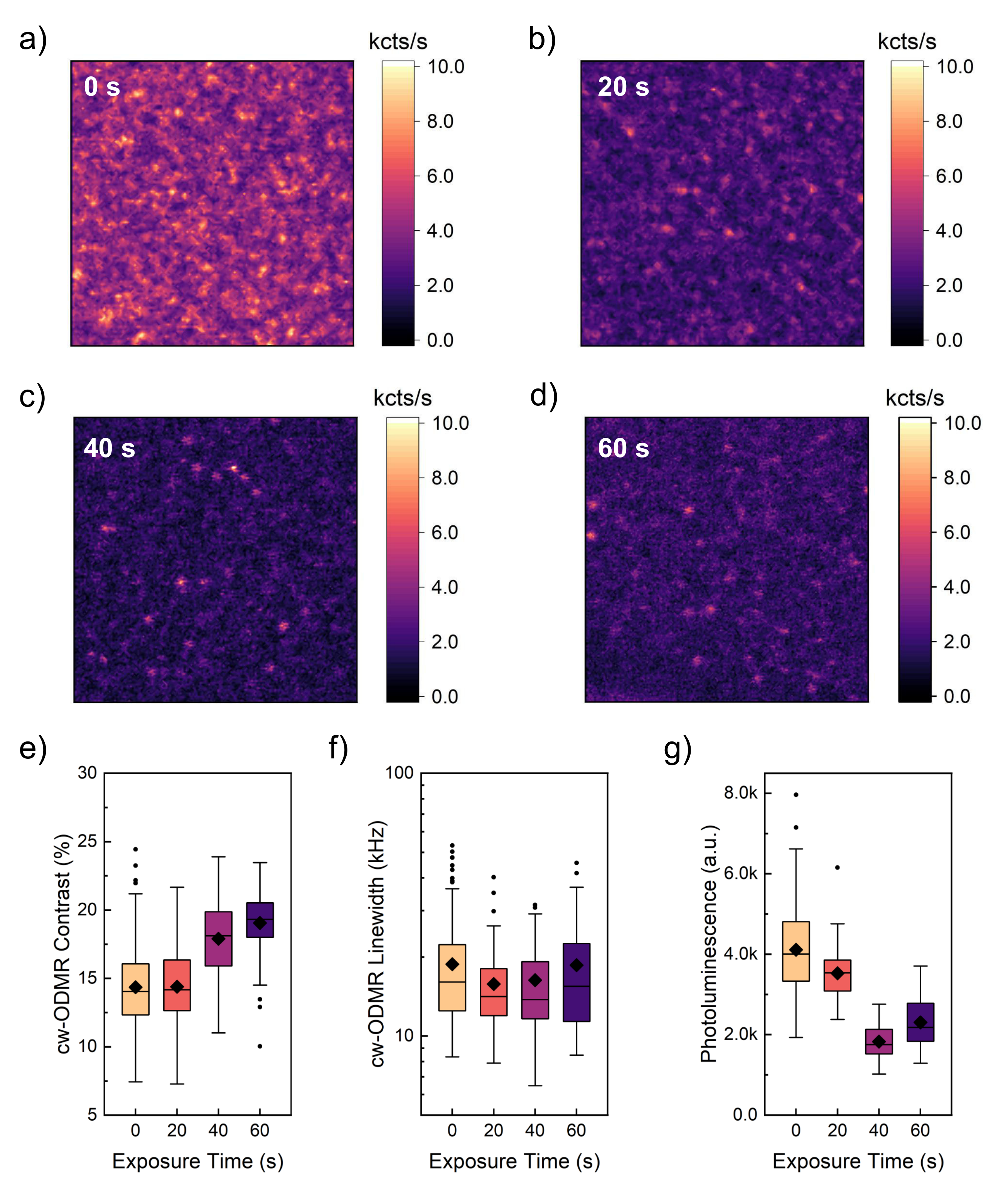}
\caption{Comparison of NV center properties in bulk diamond samples. \textbf{a)} Representative confocal fluorescence images reporting counts per second of a $20\times20$ $\si{\micro m}$ region before exposure to NH$_3$ plasma, and after \textbf{b)} 20 s, \textbf{c)} 40 s or \textbf{d)} 60 s exposure. \textbf{e)} cw-ODMR contrast, \textbf{f)} cw-ODMR linewidth, and \textbf{g)} photoluminescence intensity of NV centers before and after plasma treatment. In the box and whisker plots, black diamonds show the mean value and center lines indicate the median value of the data. Boxes compose the middle 25\%-75\% of the data, and whiskers extend to 1.5$\times$ this interquartile range. Data points shown outside of this range are outliers. \textit{N}=128, 66, 38, and 63 for 0, 20, 40, and 60 s, respectively.}
\label{fig:bulk}
\end{center}
\end{figure}

\begin{figure}[H]
\centering
\includegraphics[width=\textwidth]{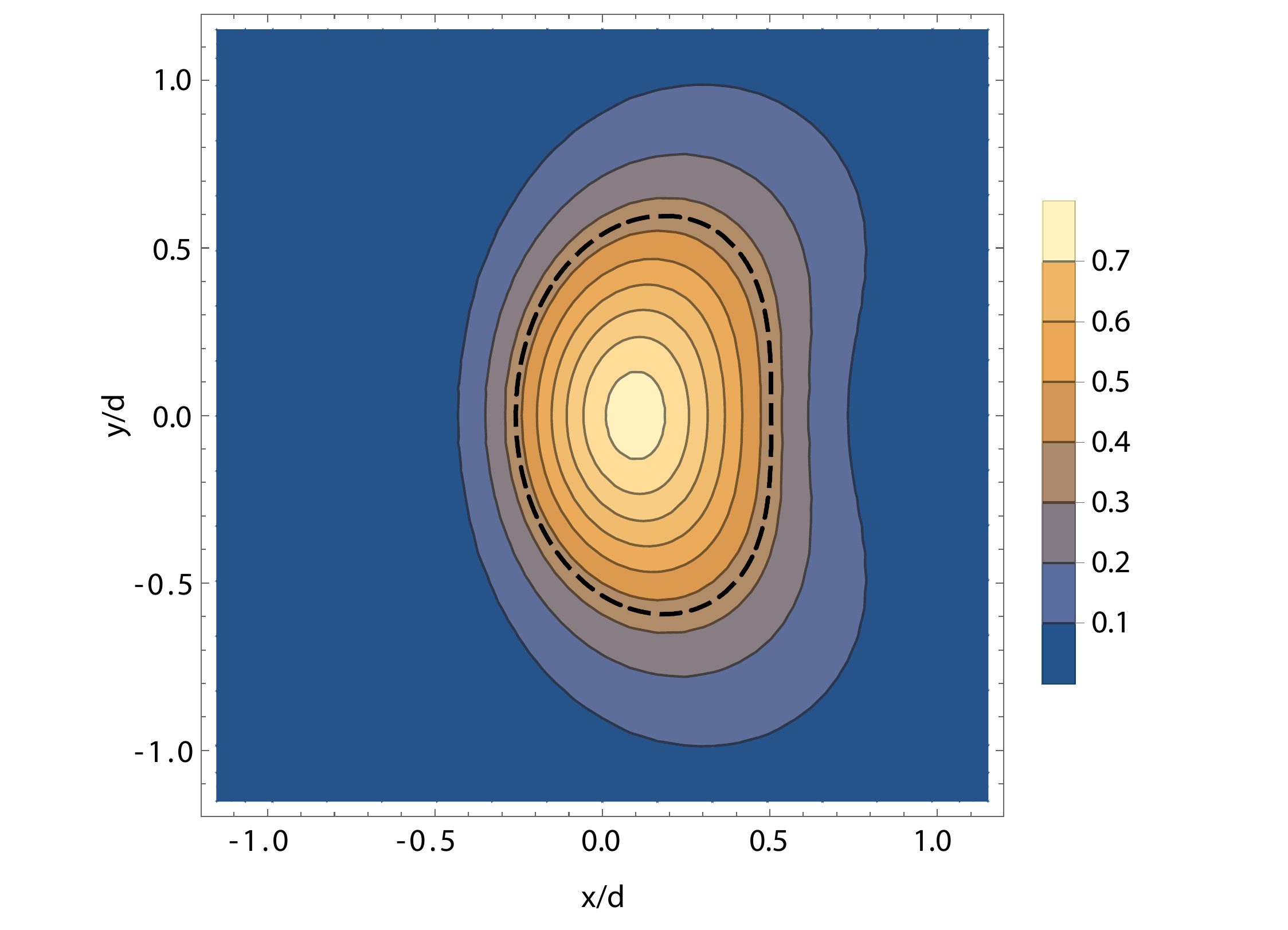}
\caption{Sensitivity region for the 2D case. Plot of $B_\mathrm{rms}^2$ in arbitrary units on the diamond surface. The NV sits at (0,0) and $d$ below the surface plane (plotting plane). The area enclosed by the dashed black path denotes the region that contributes $70\,\%$ to the $B_\mathrm{rms}$ signal.}
\label{fig:sensing_profile}
\end{figure}

\begin{figure}[H]
\centering
\includegraphics[width=\textwidth]{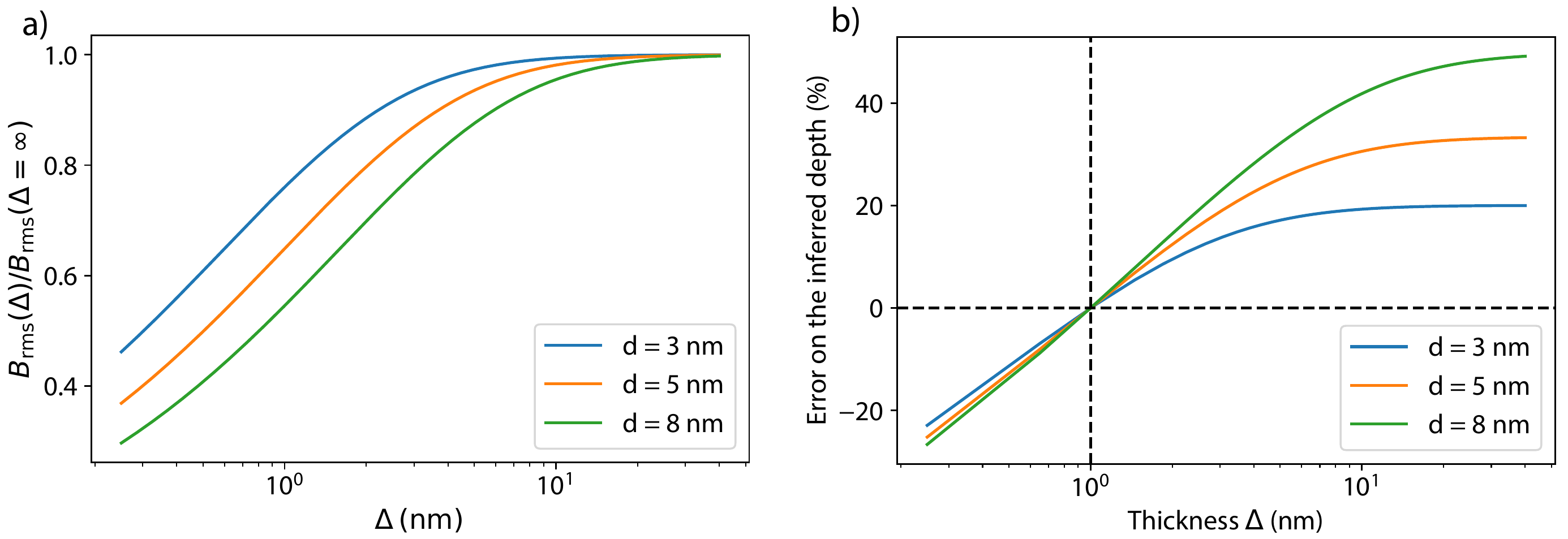}
\caption{Scaling of the estimated $B_\mathrm{rms}$ as a function of the assumed surface layer thickness for various NV depths. \textbf{a)} Square root of the under-braced factor in \eqref{eq:Brmslayer}. \textbf{b)} Error of the estimated depth for NVs when varying the assumed adsorbate layer thickness. We calculate the expected $B_\mathrm{rms}$ for NVs that are 3, 5 and 8 nm deep when a 1-nm-thick water-layer is present. Then, we invert \eqref{eq:Brmslayer} to find the depth again and vary $\Delta$. We plot the relative deviation of the inferred depth as a function of $\Delta$ compared to the "true" depth.}
\label{fig:scaling_thickness}
\end{figure}
\pagebreak
\begin{table}[]
    \centering
    \begin{tabular}{l c||c|c|c}
        { }& Thickness $\Delta$ & Depth $d$ & Density $\rho_\mathrm{2D}$ & \# of molecules\\
        \cline{2-5}
         NV 4 & 1 nm & 6.4 nm & 1.7 $\mathrm{nm}^{-2}$ & 51 \\
         { } & 2 nm & 7.3 nm & 2.9 $\mathrm{nm}^{-2}$ & 114 \\
         { } & 0.5 nm & 5.3 nm & 0.8 $\mathrm{nm}^{-2}$ & 17 \\
         { } & 0.25 nm & 4.3 nm & 0.3 $\mathrm{nm}^{-2}$ & 4 \\
         { } & $\infty$ nm & 9.1 nm & 6.9 $\mathrm{nm}^{-2}$ & 420 \\
         \cline{2-5}
         NV 6 & 1 nm & 6.3 nm & 4.0 $\mathrm{nm}^{-2}$ & 117 \\
         { } & 2 nm & 7.2 nm & 6.9 $\mathrm{nm}^{-2}$ & 263 \\
         { } & 0.5 nm & 5.5 nm & 2.3 $\mathrm{nm}^{-2}$ & 51 \\
         { } & 0.25 nm & 4.7 nm & 1.3 $\mathrm{nm}^{-2}$ & 21 \\
         { } & $\infty$ nm & 8.9 nm & 16 $\mathrm{nm}^{-2}$ & 932 \\
         \cline{2-5}
    \end{tabular}
    \caption{Estimates for the NV depth and molecular density on the surface for various thickness levels of the adsorbate layer. For NV 4, we calculated the average of the three independent $B_\mathrm{rms}$ measurements.}
    \label{tab:thicknessvariation}
\end{table}

\pagebreak
\bibliography{db}